\documentclass[twocolumn,showpacs,preprintnumbers,amsmath,amssymb]{revtex4}
\usepackage{graphicx}
\usepackage{dcolumn}
\usepackage{bm,color}
\usepackage{epsfig}
\newcommand{\rpi}{\langle r_\pi \rangle}
\newcommand{\rK}{\langle r_K \rangle}

\begin{document}

\title{Pion and kaon elastic form factors in a refined light-front model}

\author{Edson~O.~da~Silva, J.~P.~B.~C.~de~Melo, Bruno~El-Bennich and Victo~S.~Filho}
\affiliation{ Laborat\'orio de F\'{\i}sica Te\'orica e Computacional (LFTC), \\  Universidade Cruzeiro do Sul, 01506-000, S\~ao Paulo, Brazil}

\date{\today}

\begin{abstract}
Within the framework of light-front field theory, we reassess the electromagnetic form factors of the pion and kaon. Comparison with experiment
is made for the full range of momentum transfer, $q^2<0$,  including recent data. The light-front model's single regulator mass, $m_R$, of the $\bar qq$
bound-state vertex function is initially adjusted to reproduce the weak decay constants, $f_\pi$ and $f_K$, and both meson's charge radii, $\rpi$ and $\rK$. 
We study the behavior of these observables under variation of the quark masses and find an optimized parameter set, $m_u=m_d$, $m_s$ and $m_R$, 
for which they are in sensibly better agreement with experiment than in a previous analysis; a feature also observed for the elastic form factors, in particular 
at small $q^2$. This model refinement is important in view of an extension to vector and heavy-light mesons. 

\end{abstract}

\maketitle

\noindent
\textbf{\em Introduction }
The light pseudoscalar mesons play a crucial role in the understanding of low-energy QCD, being the lightest strongly bound antiquark-quark states as well as the Goldstone
bosons associated with chiral symmetry breaking. Their static properties have been extensively studied---see, {\em e.g.\/}, Refs.~\cite{Tobias92,Maris:1997hd,Maris:1997tm,Choi:1998nf,
Maris:1999nt,Maris:1999bh,Ebert:2005es,Bakker:2000pk}---while dynamical features have also been investigated theoretically~\cite{Lepage:1979zb,Lepage:1980fj,
Frederico:1994dx,Maris:2000sk,Maris:2002mz,Roberts:2010rn,Nguyen:2011jy,Simula2002,Pacheco99,Pacheco2002,deMelo:2003uk,deMelo:2005cy,deMelo:2006rg,
ElBennich:2008qa,ElBennich:2004px,ElBennich:2004jb,Brodsky:2007hb,Loewe2007,Krutov:2009tm,Leitner2011,Choi:1998jd} and experimentally \cite{Brauel,Data3,Data2a,
Tadevosyan,Horn,Huber}. With respect to the description of bound states on the light cone, a detailed review of hadronic wave functions in QCD models can be found 
in Ref.~\cite{Brodsky}.

In the case of the pion, a great deal has been learned experimentally from its electromagnetic form factor $F_\pi (q^2)$~\cite{Data3,Data2a,Huber,Tadevosyan,Brauel,Horn},
whereas this is not the case for the kaon~\cite{Amendolia:1986ui,Dally:1980dj}. Additional important knowledge about the mesons' internal structure can be inferred from their 
valence-quark parton distribution functions~\cite{Nguyen:2011jy}. 

The theoretical framework we adopt is the light-front field theory formalism~\cite{Brodsky}. More specifically, we here ameliorate the light-front approach first introduced 
in Refs.~\cite{Tobias92,Frederico:1994dx}, where two classes of $\bar qq$ bound-state models for the Bethe-Salpeter amplitude of the pion must be distinguished: 
the nonsymmetric~\cite{Pacheco99} and the symmetric~\cite{Pacheco2002} vertex model. The light-front component $J^+$ of the electromagnetic current has been 
successfully used  to calculate elastic form factors~\cite{Dziembowski87,Tobias92,Cardarelli96,Simula2002,Jaus99,Hwang2003,Huang2004}. For the nonsymmetric
$\pi\bar qq$ and $K\bar qq$ vertex models, the components of the current are conveniently obtained in the Drell-Yan frame. We recall that on the light cone the bound 
state wavefunctions are defined on the hypersurface $x^0+x^3=0$ and are covariant under kinematical boosts due to the stability of Fock-state decomposition~\cite{Perry90}. 

In this Brief Report, we solely consider the nonsymmetrical $\bar qq$ bound-state vertex function with the intention to optimize and unify the parameter set which simultaneously 
reproduces the pion and kaon decay constants, charge radii and their electromagnetic form factors. For the latter, our numerical results are compared with experimental 
data up to 10~GeV$^2$ in order to explore the validity of the model at large $q^2$ transfer. 

\noindent
\textbf{\em The model } We briefly summarize the light-front model including the {\em Ansatz\/} for the nonsymmetric vertex function for the pseudoscalar bound states.
The covariant electromagnetic form factor is defined as (suppressing color and flavor indices), 
\begin{equation}
   (p+p^{\prime})_\mu \, F_{M_{0^-}}(q^2) \ = \ \langle M_{0^-}(p^{\prime}) | J_\mu | M_{0^-} (p) \rangle \ , 
\label{ffactor}
\end{equation}
in which $M_{0^-} = \pi, K$ denotes the light pseudoscalar mesons, $q =p- p^\prime$ and $J_{\mu} =  e_q\,\bar \psi_q \gamma_\mu \psi_q$ is the electromagnetic current.
In the impulse approximation, the form factor is given by a triangle diagram which represents an amplitude via a single integral:
\begin{eqnarray}
 \lefteqn{  \hspace*{-4mm}
 (p_\mu+p^\prime_ \mu)F_{M_{0^-}}(q^2)   =   e_q N_c  \int\! \frac{d^4k}{(2\pi)^4}\,  \Gamma(k,p')\Gamma(k,p)   } \nonumber \\ 
    & &  \hspace*{-5mm}\times\, \mathrm{Tr} \Bigl[ S_{\bar q}(k) \gamma_5 S_q(k-p^{\prime})  \gamma_\mu S_q(k-p) \gamma_5 \Bigr ] + [ q \leftrightarrow  \bar q ]\,  , 
\label{current}
\end{eqnarray} 
where $S^{-1}_{q}(p)=\ \rlap\slash p-m_q+\imath \epsilon$, $q=u$, $\bar q= \bar d, \bar s$, denotes the quark propagator with  quark masses $m_q$,
$e_q$ is the quark's charge and $N_c=3$ is the number of colors. The bracket $[q \leftrightarrow \bar q\, ]$ is a shorthand for the exchange of the quark and antiquark
in the integral. In the following, we work in the Breit frame with the Drell-Yan condition, such that $p_{\mu}=(p_0,q/2,0,0)$, $p^\prime_\mu=(p_0',-q/2,0,0)$, $p_0=p_0'$ 
and $q_\mu=(0,q,0,0)$. 

After transformation to light-cone variables, using the plus component, $J^+$, of the electromagnetic current, the first integration is over $k^-$ (the null-plane energy) 
whose pole contribution is $\bar k^- = (k_\perp+m_q-\imath \epsilon)/k^+$. It was shown that $k^+$ is limited to two integration intervals: the valence contribution, constrained by 
$0 < k^+ < p^+$, while the nonvalence contribution is restricted to $p^+ < k^+ < p'^{+}$~\cite{Pacheco99,Pacheco2002}. Thus, on the light front, two distinct terms contribute 
to the electromagnetic form factor: 
\begin{eqnarray}  
   F_{M_{0^-}}(q^2) = F^{\mathrm{I}}_{M_{0^-}}(q^2)+F^{\mathrm{II}}_{M_{0^-}}(q^2), 
\label{ffactor1}
\end{eqnarray}
where $F^{\mathrm{I}}_{M_{0^-}}(q^2)$ and $F^{\mathrm{II}}_{M_{0^-}}(q^2)$ are the  valence $(\bar qq)$ and nonvalence (pair production) contributions, respectively. 

Furthermore, it can be shown that for a non symmetric bound-state vertex function, $\Gamma (k,p)$, asymmetric under momentum exchange of the quark and antiquark, the second interval 
does not contribute to the electromagnetic form factor~\cite{Pacheco99}. Within a quark-meson interaction model with effective coupling,
\begin{equation}
  \mathcal{L}  = -\imath\, \frac{\hat m}{f_{M_{0^-}}} \, \vec \pi \cdot \bar q\, \gamma_5 \vec \tau q \ ,
\end{equation}
\\
a nonsymmetric vertex function is 
\begin{equation}
   \Gamma (k,p)  =   \frac{N}{(p-k)^2-m^2_R+\imath\epsilon} \ ,
\label{nosymm} 
\end{equation}
where,  $\hat m = (m_q +m_{\bar q})/2$,  $f_{M_{0^-}}$ is the weak decay constant, $m_R$ is the regulator mass and $N$ is an overall normalization. 
Thus, after $k^-$ integration and change of variables, $x=k^+/p^+$, where $x$ is the momentum fraction carried by the quark in the infinite-momentum frame, the elastic form
 factor is given by the integral ($F^+_{M_{0^-}} =  F^+_{M_{0^-}} (q^2)$):
\begin{eqnarray}
\lefteqn{ F^+_{M_{0^-}} \!  =\,  \frac{e_q N_c}{2p^+}\!\! \int\! \frac{d^{2} k_{\perp} d x}{(2 \pi)^3 x }\,  \Psi^*\!(x,k_{\perp})  \Psi (x,k_{\perp}) \theta(1-x)  \theta(x) }  \nonumber  \\
 & &   \mathrm{Tr}\! \left [ (\rlap\slash k +m_{\bar{q}}) \gamma^5 (\rlap\slash k-\rlap\slash p^{\prime}+m_q) \gamma^+ (\rlap\slash k-\rlap\slash p +m_q) 
                                    \gamma^5 \right ]_{k^-=\bar k^-}  \nonumber  \\
  &  & \ \ + \ [\, q \leftrightarrow  \bar q \,  ] \ .
\label{form}
\end{eqnarray}
The subscript ``$+$"  on the form factor in Eq.~\eqref{form} is a reminder that we employ the $J^+$ component of the electromagnetic current. For $M_{0^-}=\pi^+$, the quark
flavors are $q=u$ and $\bar q = \bar d$, whereas for $M_{0^-}=K^+$ we have $q=u$ and $\bar q = \bar s$ and $[ q \leftrightarrow  \bar q   ]$ is again an abbreviation for both quark
and antiquark contributions to the elastic form factor. Moreover, after $k^-$ integration, a light-front wave function emerges which for a nonsymmetric $\bar qq$ vertex function
is defined as:
\begin{equation}
\Psi (x,k_{\perp}) =  \frac{\hat m}{f_{M_{0^-}}} \frac{N}{(1-x)^2 (m_{M_{0^-}}^2\! - {\cal M}_0^2) (m_{M_{0^-}}^2\! - {\cal M}_R^2)}  \  ,
\label{wavefunction}
\end{equation}
where ${\cal M}_R={\cal M} (m_q^2,m^2_R)$ is given by,
\begin{equation}
  {\cal M}_R^2 = \frac{k_{\perp}^2+m_q^2}{x} + \frac{(p-k)_{\perp}^2+m_R^2}{(1-x)}-p^2_{\perp} \ ,
\end{equation} 
${\cal M}^2_0={\cal M}^2(m_q^2,m_q^2)$ is the free mass operator, $m_{M_{0^-}}$ is the pseudoscalar meson mass and the normalization constant $N$ obeys the 
condition $F^+_{M_{0^-}}(0)=1$. 

In addition, we calculate the pseudoscalar weak decay constant with the same bound-state vertex model introduced in Eq.~\eqref{nosymm},
\begin{equation}
\langle 0 | A_\mu (0) |{M_{0^-}} (p) \rangle =  \imath\, f_{M_{0^-}} p_\mu \ ,
\end{equation}
where $A_\mu = \imath\, \bar \psi_q \gamma_\mu \gamma_5 \psi_q$.
After Dirac algebra and $k^-$ integration, the pseudoscalar decay constant reads,
\begin{equation}
 \hspace*{-2mm}  f_{M_{0^-}} \! = \, N_c \! \int  \frac{d^2k_\perp dx}{(2 \pi)^3 x} \, \big [ 4 x m_{\bar q } + 4  m_{q} (1-x)\big ]  \Psi (x,k_\perp) .
\end{equation}
We remind that the charge radius is determined via,
\begin{equation}
   \langle\, r^2_{M_{0^-}} \rangle = -6 \left [ \frac{  dF_{M_{0^-}} (q^2) }{dq^2} \right ]_{q^2=0} .
\end{equation}

\noindent
\textbf{\em Numerical Results }  We have three model parameters: the regulator mass, $m_R$, and the quark masses, $m_u =m_d$ and $m_s$. The main aim of this work 
is to {\em jointly\/} analyze the pion's and kaon's elastic form factors, decay constants and charge radii in order to determine more accurately the model's quark masses in view 
of future applications and to test whether a single mass scale, $m_R$, can satisfactorily describe experimental data for both light mesons.

We initially consider the parameters of Ref.~\cite{LC09}, {\em i.e.\/} $m_u=m_d=220$~MeV, $m_s =419$~MeV, $ m_{\pi^+} = 140$~MeV, $ m_{K^+} = 494$~MeV and 
$m_R =  0.946$~GeV, in which the elastic form factor ratio of the kaon and pion were calculated. This parameter set describes rather well measured observables for the pion 
but less so for the kaon: $\langle r_{\pi^+} \rangle= \surd \langle r_{\pi^+}^2  \rangle =  0.672$~fm, which coincides with the experimental value, $\langle \, r^{\mathrm{exp.}}_{\pi^+} \rangle 
= 0.672\pm0.008$~fm~\cite{Amendolia:1986wj}, and $f_\pi = 101$~MeV ($f_\pi^{\mathrm{exp.}}/\surd{2}=92.21\pm 0.14$~MeV~\cite{PDG2012}); whereas 
$\langle r_{K^+} \rangle = \surd \langle r_{K^+}^2 \rangle = 0.71$~fm ($\langle \, r^{\mathrm{exp.}}_{K^+} \rangle = 0.560\pm 0.031$~fm~\cite{Amendolia:1986ui}) and 
$f_K = 129$~MeV ($f_K^{\mathrm{exp.}}/\surd{2} =  110.4\pm 0.6$~MeV~\cite{PDG2012}) indicate that the model is not well adjusted for mesons with strangeness content.

We vary the masses of the constituent quarks and simultaneously tune $m_R$ toward a common value for both the pion and the kaon. Clearly, the 
regulator mass acts as a cut-off in the triangle diagram but also defines a physical length scale. Therefore, one would not expect $m_R$ to be equal for both mesons. 
However, we here insist on having a minimal number of parameters which still yields a satisfying reproduction of all available data. Once the light quark masses 
in the model are fixed, we can consider heavier mesons as well as $1^-$ mesons~\cite{pacheco2011,pacheco2012} and compute their 
electromagnetic properties~\cite{pachecofuture}.

\begin{figure}[t]
\vspace*{-3mm}
\includegraphics[scale=.32]{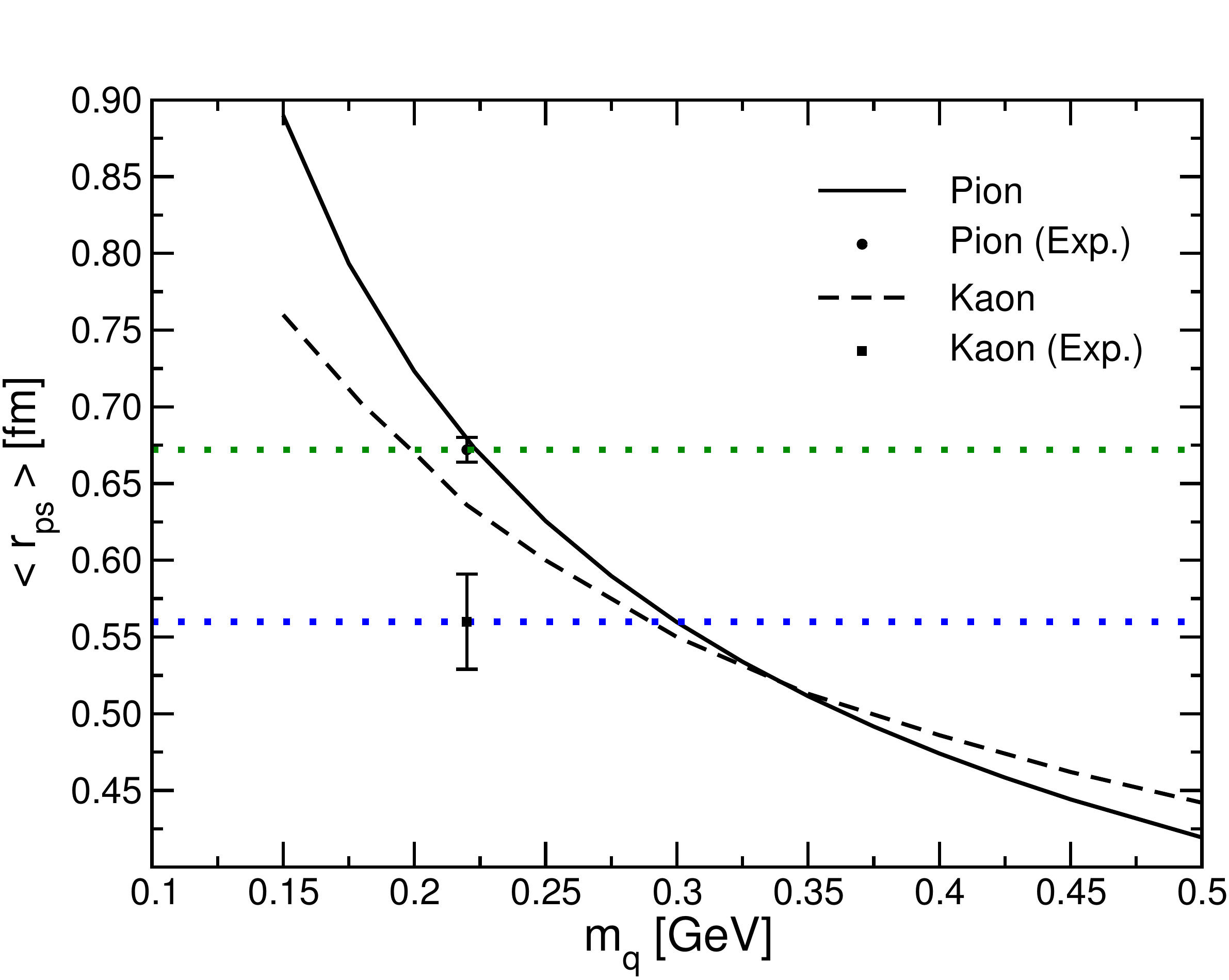}
\caption{Pion and kaon charge radii, $\sqrt{\langle r^2 \rangle}$, as a function of $m_u=m_d$; $m_s=0.51$~GeV and $m_R=1.0$~GeV are fixed. 
Upper and lower dotted lines mark experimental values for the pion and kaon, respectively.}
\label{lcp1}
\vspace*{-1mm}
\end{figure}

\begin{figure}[t]
\includegraphics[scale=.32]{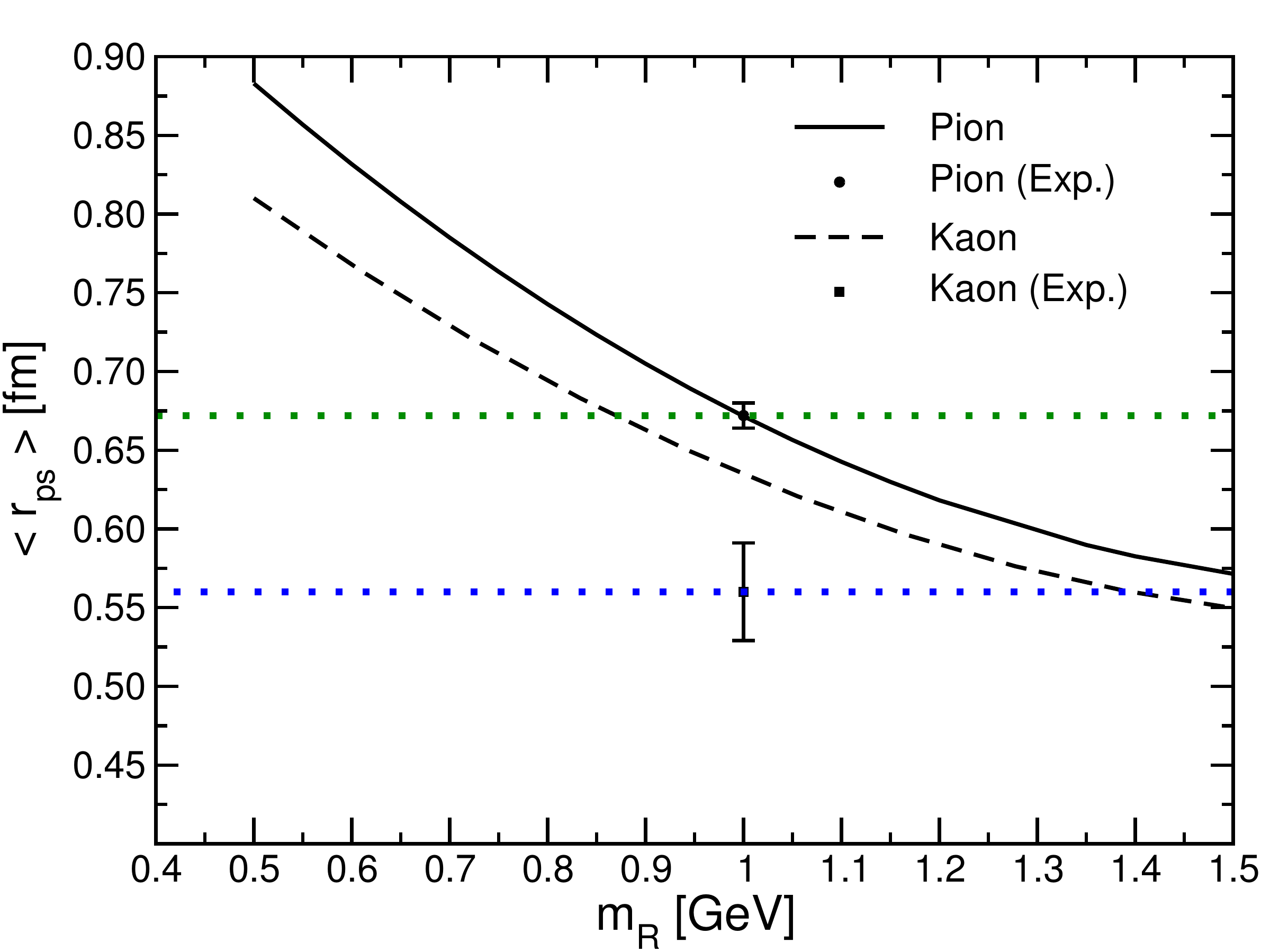}
\caption{Charge radii, $\sqrt{\langle r^2 \rangle}$, of the pion and kaon as function of $m_R$ with $m_u=m_d=0.22$~GeV and $m_s=0.51$~GeV fixed.
Horizontal dotted lines as in Fig.~\ref{lcp1}. }
\label{lcp2}
\vspace*{-3mm}
\end{figure}

In Fig.~\ref{lcp1}, we plot the charge radii $\langle r_\pi \rangle$ and $\langle r_K \rangle$ as a function of $m_u=m_d$, where $m_R=1.0$~GeV and $m_s=510$~MeV 
are kept fixed, whereas in Fig.~\ref{lcp2} the functional behavior of the charge radii in dependence of  $m_R$ is shown with $m_u=220$~MeV and $m_s=510$~MeV fixed. 
The dependence is in both cases nonlinear and somewhat more pronounced for variations of the quark mass than of the regulator mass. We remark that a larger regulator 
mass results in a smaller charge radius, as expected. The strange constituent quark mass has been readjusted from its value $m_s=451$~MeV~\cite{LC09} to obtain a 
better agreement with the kaon charge radius while keeping $m_R=1.0$~GeV. The value $m_s = 510$~MeV we here choose is in agreement with a definition of the 
Euclidean constituent quark mass derived from solutions of Dyson-Schwinger equations~\cite{ElBennich:2012tp,Bashir:2012fs}.

\begin{figure}[t]
\vspace*{-3mm}
\includegraphics[scale=.32]{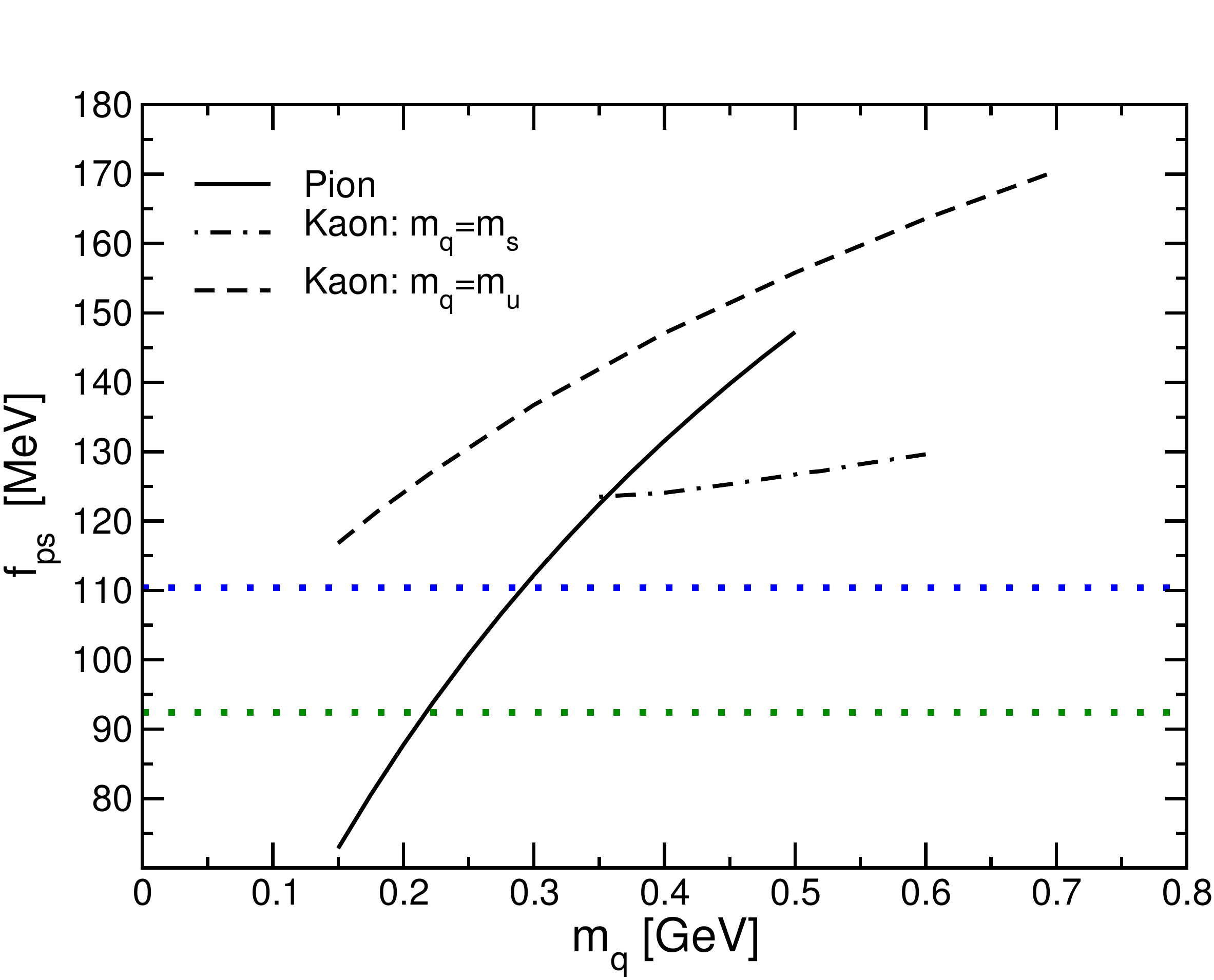}
\caption{Weak decay constants: $f_\pi$ as function of $m_u=m_d$ (solid line) and $f_K$ as function of $m_u$ (dashed line with $m_s=0.51$~GeV) 
and $m_s$ (dot-dashed line with $m_u=0.22$ GeV); $m_R=1.0$ GeV in all cases. The upper and lower dotted lines denote experimental values for $f_K$ and $f_\pi$, respectively.}
\label{lcp3}
\vspace*{-4mm}
\end{figure}

\begin{figure}[t]
\includegraphics[scale=.32]{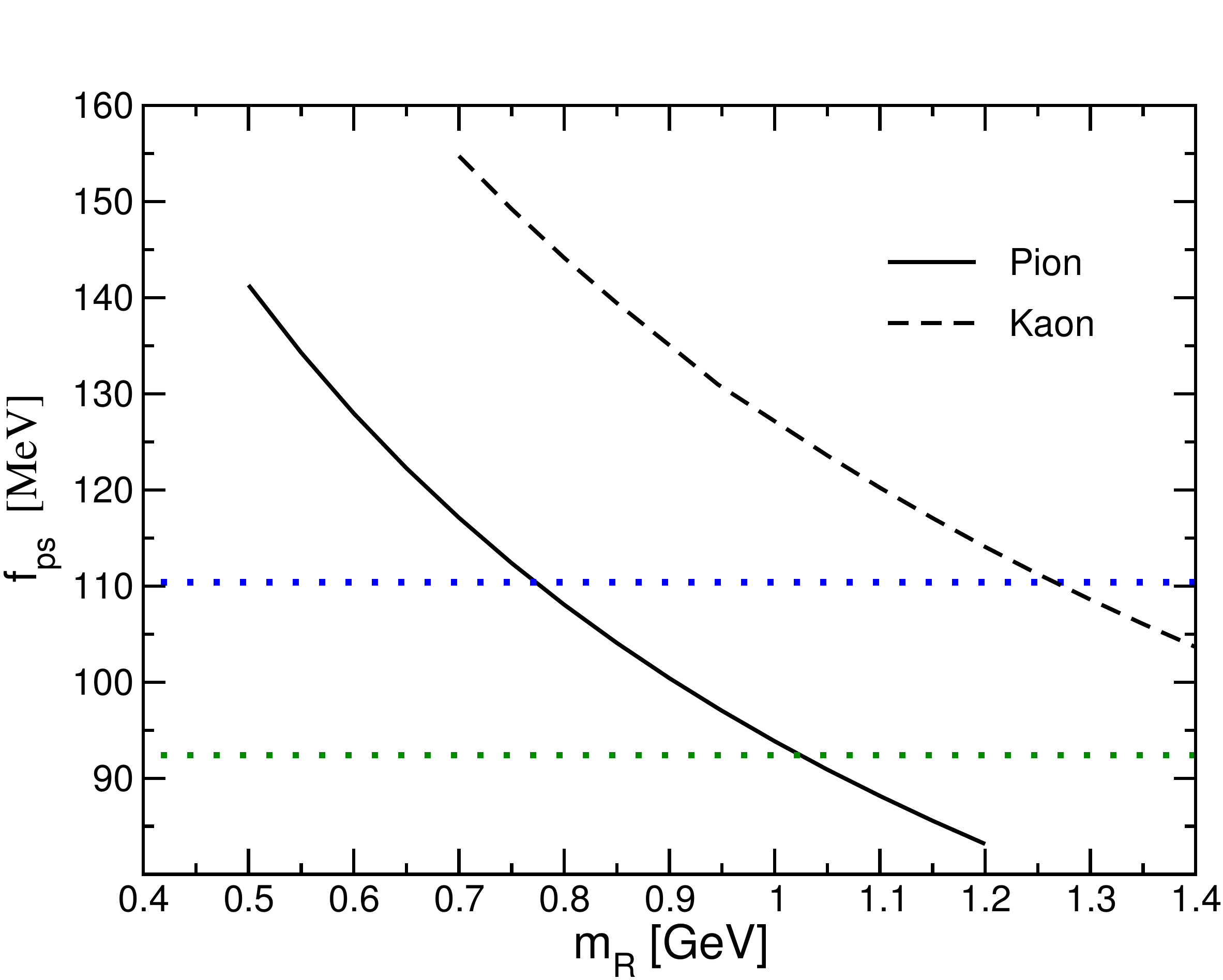}
\caption{The weak decay constants, $f_\pi$ and $f_K$, as a function of the regulator mass, $m_R$, with $m_u=m_d=0.22$~GeV and $m_s=0.51$~GeV fixed. 
Horizontal dotted lines as in Fig.~\ref{lcp3}.}
\label{lcp4}
\vspace*{-2mm}
\end{figure}

\begin{figure}[t]
\vspace*{-6mm}
\includegraphics[angle=0, scale=.33]{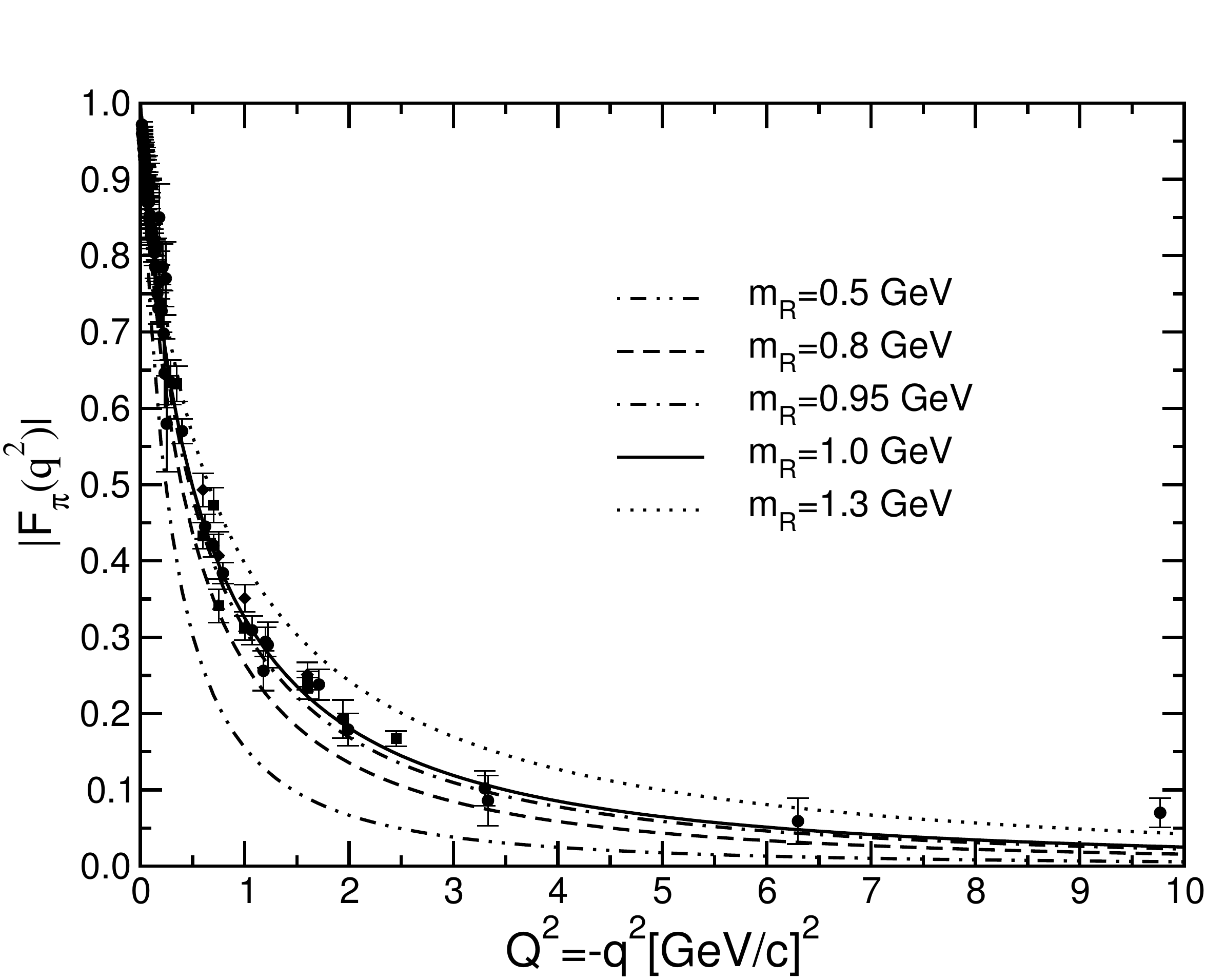}
\caption{Elastic form factor of the pion as function of $m_{R}$ and with $m_u=m_d=0.22$~GeV. 
Experimental data:  Refs.~\cite{Data3,Brauel} (solid triangles), Ref.~\cite{Horn} (solid diamonds), Ref.~\cite{Tadevosyan} (solid circles) and Ref.~\cite{Huber} (solid squares).  }
\label{lcp5}
\end{figure}

\begin{figure}[t]
\includegraphics[angle=0, scale=.33]{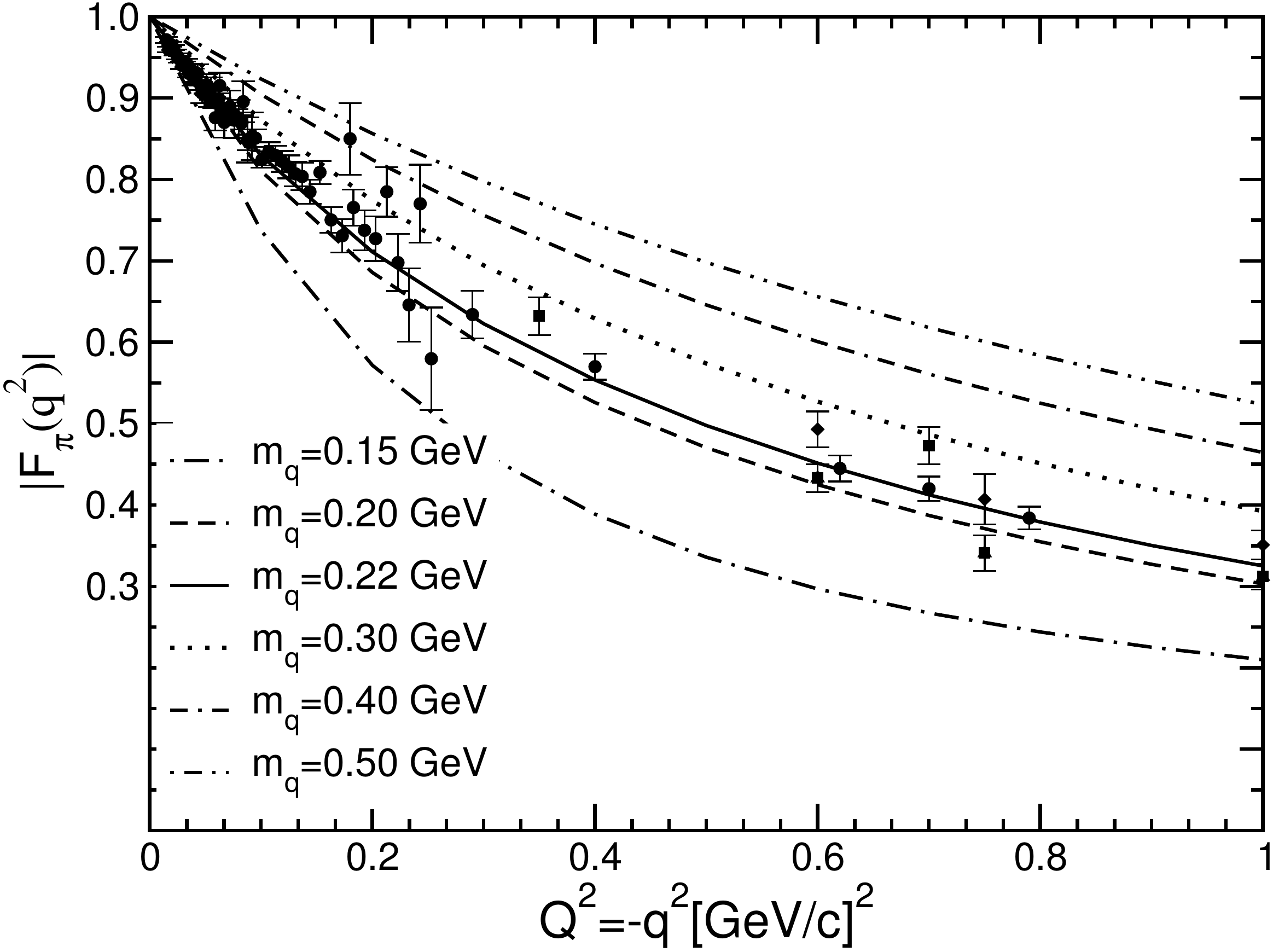}
\caption{Elastic form factor of the pion as function of $m_{u,d}$ = 0.22 GeV and $m_R=1.0$~GeV. 
Experimental data as in Fig.~\ref{lcp5}. }
\label{lcp7}
\end{figure}

Similarly, in Fig.~\ref{lcp3} the decay constants $f_\pi$ and $f_K$ are shown as a function of the quark mass ($m_R = 1.0$~GeV) and in Fig.~\ref{lcp4} as a function of 
the regulator mass (where $m_{u,d}=220$~MeV and $m_s=510$~MeV). We note that the decay constants are also more sensitive to variations of $m_q$ than of $m_R$,
excepting in the case of the strange quark where a departure of about 10\% from $m_s=510$~MeV does not significantly alter $f_K$. For the pion, $m_R=1.0$~GeV yields 
the best adjustment to the experimental values of the decay constant and charge radius; since it also occurs to best describe the pion's elastic form factor in  Figs.~\ref{lcp5} and 
\ref{lcp7}, as discussed below, we definitely set $m_R=1.0$~GeV and use this value in calculations of the kaon's properties. Moreover, while $\langle r_\pi\rangle$ and $\langle r_K\rangle$ 
decrease with increasing quark mass, the opposite is true for $f_\pi$ and $f_K$, which is an expression of the Tarrach relation, $\langle r_{M_{0^-}} \rangle\sim 1/m_q$ and 
$f_{M_{0^-}} \sim 1/\langle r_{M_{0^-}} \rangle$~\cite{Tarrach:1979ta}.

\begin{figure}[t]
\includegraphics[angle=0, scale=.33]{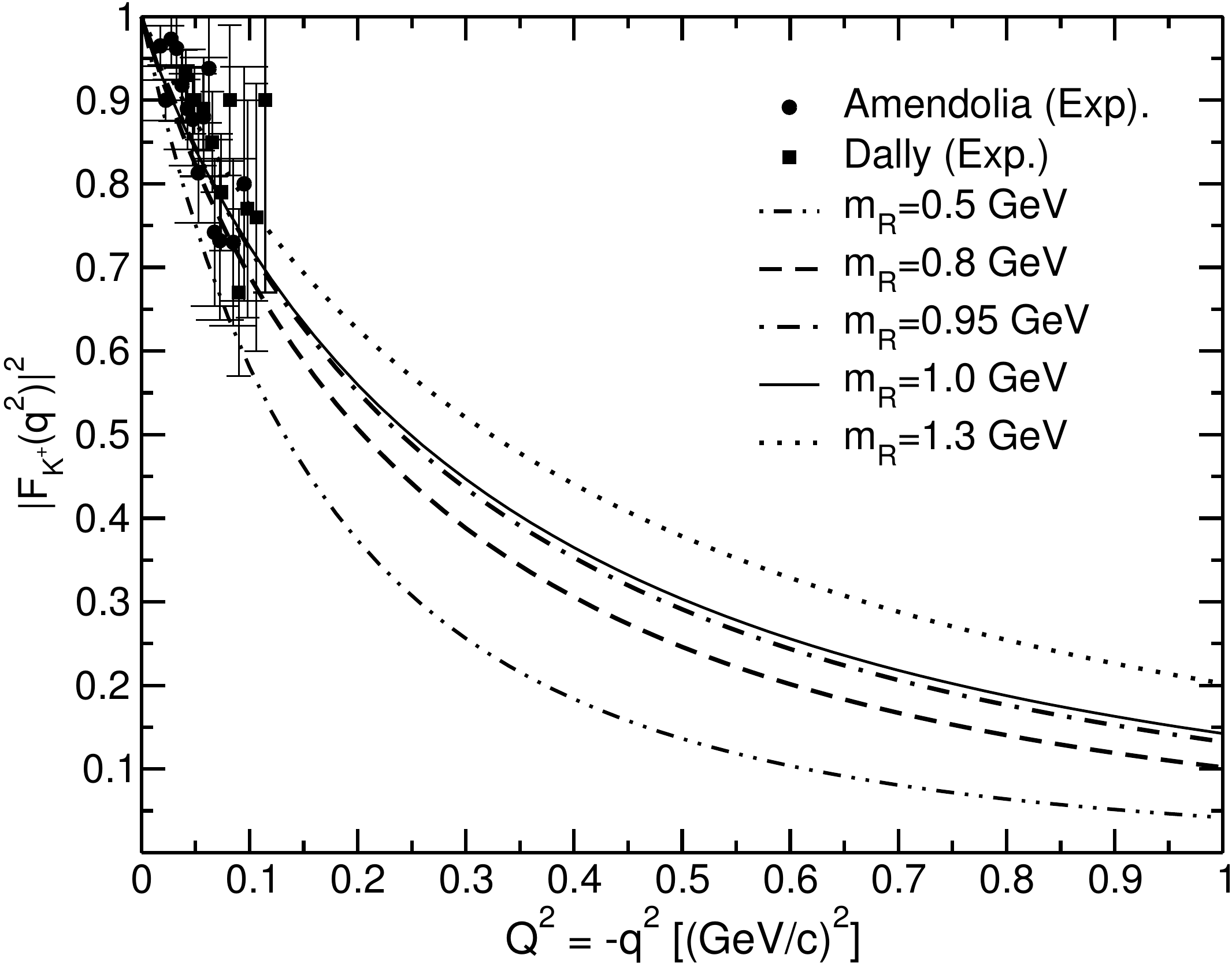}   
\caption{Elastic form factor of the kaon  as function of  $m_{R}$  with~$m_u=m_d=0.22$~GeV 
and $m_{\bar{s}}=0.51$~GeV; experimental data from Refs.~\cite{Amendolia:1986ui,Dally:1980dj}.}
\label{lcp6}
\vspace*{-3mm}
\end{figure}

We thus require $m_R=1.0$~GeV, $m_u=m_d =220$~MeV and $m_s=0.51$~GeV as our reference values and observe their implications for both pseudoscalars' elastic form 
factors. In case of the pion, this mass choice in the light-front model reproduces very well the experimental data ~\cite{Huber,Tadevosyan,Data3,Brauel,Horn},
while the lack of data on the kaon's elastic form factor and associated large error bars do not allow for a satisfactory comparison. 

For the sake of completeness, we illustrate the strong sensitivity of the elastic form factors to $m_R$. As can be seen from Figs.~\ref{lcp5} and \ref{lcp6}, acceptable values  
with respect to  the experimental data on $F_\pi(q^2)$ and $F_K (q^2)$ lie in the interval, $0.8$~GeV $\lesssim m_R \lesssim 1.3$~GeV, which coincides with our privileged value 
$m_R=1.0$~GeV. 

\begin{figure}[t]
\includegraphics[angle=0, scale=.33]{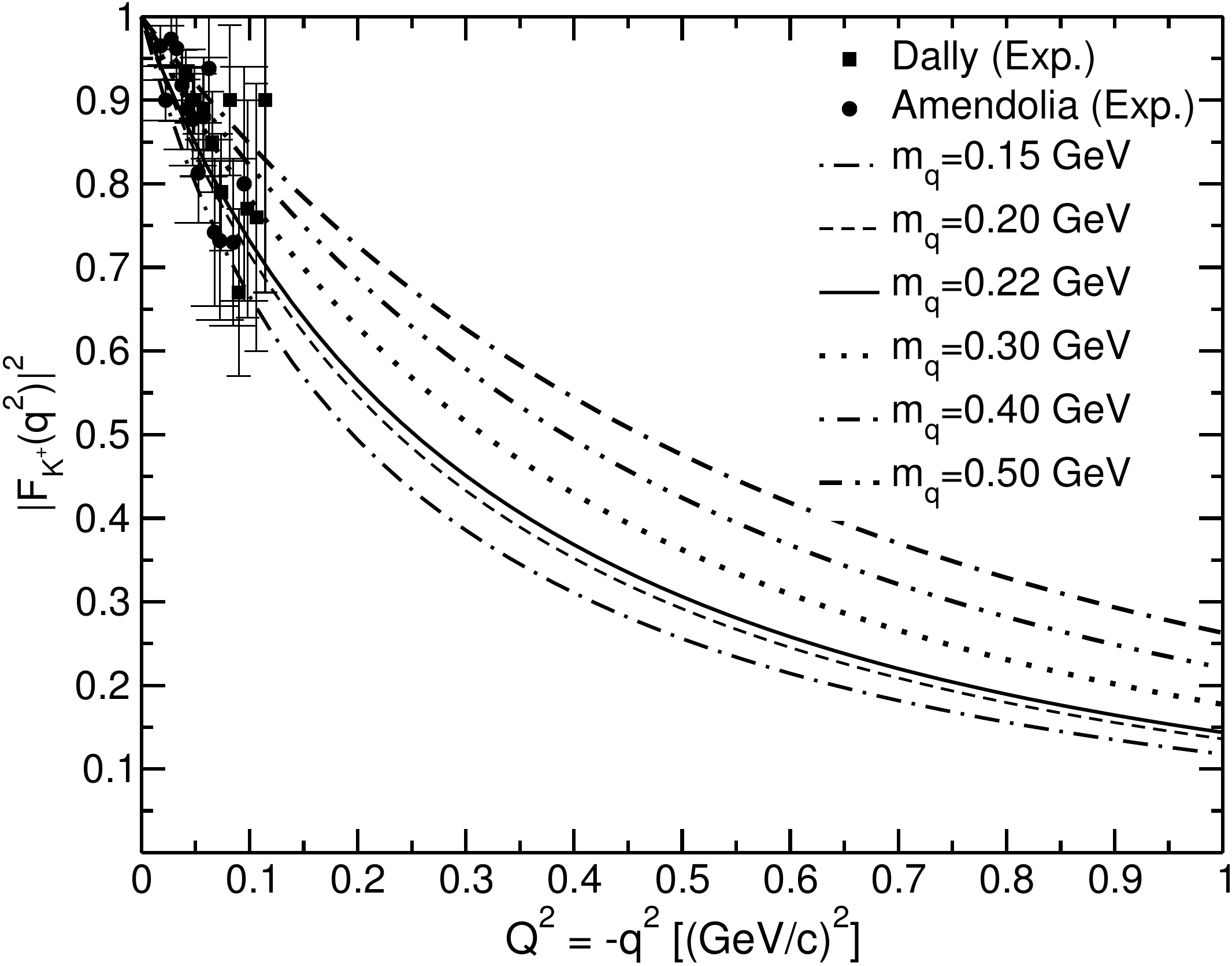}  
\caption{Elastic form factor of the kaon in dependence on the quark mass, $m_u$, where $m_R=1.0$~GeV and $m_{\bar{s}}=0.508$~GeV 
are fixed; experimental data are from Refs.~\cite{Amendolia:1986ui,Dally:1980dj}.}
\label{lcp8}
\vspace*{-4mm}
\end{figure}

Next, the model dependence on the constituent masses are explored in Figs.~\ref{lcp7} and \ref{lcp8}. For the pion, we observe that this dependence 
is asymmetric and more strongly pronounced for smaller quark-mass values. For instance, compared with the reference mass $m_u = 220$~MeV, a $70$~MeV 
lighter constituent quark yields a soft elastic form factor that strikingly deviates from the experimental data, whereas the same calculation with  $m_u = 300$~MeV results 
in just a slightly harder form factor, in particular for $Q^2=-q^2 \lesssim 0.2$~GeV$^2$. This is not surprising, as this momentum range probes the static features of the 
mesons where a constituent quark mass of $200-350$~MeV is appropriate~\cite{GutierrezGuerrero:2010md}. For larger values of $q^2$, however, this is not the case 
and a simple model will fail. This asymmetric behavior with respect to the light-quark mass is not observed in Fig.~\ref{lcp8}, where the kaon's dynamical features are 
dominated by the strange mass.

These observations on the elastic form factors confirm our choice of mass parameters, $m_R=1.0$~GeV, $m_u=m_d =220$~MeV and $m_s=510$~MeV, 
for which we obtain:
\begin{align*}
  f_\pi & =  93.1\ \mathrm{MeV}  & (f_\pi^{\mathrm{exp.}} & =  92.42 \  \mathrm{MeV}) \\
  f_K   & = 126.9 \  \mathrm{MeV} & (f_K^{\mathrm{exp.}} & = 110.4 \ \mathrm{MeV}) \\
 \langle r_\pi \rangle & =  0.679\ \mathrm{fm}  & 
(r_\pi^{\mathrm{exp.}} & =  0.672 \  \mathrm{fm})  \\
 \langle r_K  \rangle & =  0.636 \ \mathrm{fm}  & 
(r_K^{\mathrm{exp.}} & =  0.560 \  \mathrm{fm}) \ .
  \end{align*}

\noindent
\textbf{\em Conclusions }
We reassessed the light-front model~\cite{Pacheco99,Pacheco2002} in view of recent data on $F_\pi(q^2)$ and due to the need of a better determination and restriction 
of the model parameters. The regulator mass, $m_R = 1.0$~GeV, using a nonsymmetric vertex model for the bound state, is found to simultaneously satisfy the experimental 
data on the space-like elastic form factors, the weak decay constants and the charge radii of the pion and the kaon within reasonable theoretical uncertainties. The numerical 
results show that the model significantly breaks down for $m_R \lesssim 0.8$~GeV and for $m_R \gtrsim 1.3$~GeV. We have also studied the model dependence on the 
constituent masses, $m_u=m_d$ and $m_s$, showing that Tarrach's relation is satisfied and confirming the range of mass values  commonly chosen within the light-front 
model. These parameter values are useful for, {\em e.g.\/} heavy-meson and vector decay constants~\cite{pachecofuture} or heavy-to-light transition form factors.

\noindent
\textbf{\em Acknowledgments }
E.~O.~S. acknowledges support from CAPES. J.~P.~B.~C.~M. and V.~S. F. are partially supported by FAPESP grant no.~2009/53351-0. 
B.~E. is funded by FAPESP, grant nos.~2009/51296-1 and 2010/05772-3.

\end{document}